\documentclass[sigconf]{acmart}

\usepackage{booktabs} 
\usepackage{breqn}
\usepackage{graphicx}
\graphicspath{{images/}}

\makeatletter
\g@addto@macro\normalsize{%
  \setlength\abovedisplayskip{0pt}
  \setlength\belowdisplayskip{0pt}
  \setlength\abovedisplayshortskip{-5pt}
  \setlength\belowdisplayshortskip{0pt}
  \setlength{\abovecaptionskip}{3pt}
  \setlength{\belowcaptionskip}{-15pt}
  \setlength{\parskip}{0pt}
  \setlength{\parsep}{0pt}
  \setlength{\topskip}{0pt}
  \setlength{\partopsep}{0pt}
}
\makeatother

\setcopyright{rightsretained}

\acmDOI{10.475/123_4}

\acmISBN{123-4567-24-567/08/06}

\begin{document}
\title{Analyzing Evolving Stories in News Articles}

\author{Roberto Camacho Barranco}
\orcid{0000-0002-5821-4074}
\affiliation{
  \institution{The University of Texas at El Paso}
  \streetaddress{500 W University Ave.}
  \city{El Paso} 
  \state{Texas} 
  \postcode{79968}}
\email{rcamachobarranco@utep.edu}

\author{Arnold P. Boedihardjo}
\affiliation{
  \institution{U. S. Army Corps of Engineers}
  \streetaddress{7701 Telegraph Rd}
  \city{Alexandria} 
  \state{Virginia} 
  \postcode{22315}}
\email{arnold.p.boedihardjo@usace.army.mil}

\author{M. Shahriar Hossain}
\affiliation{
  \institution{The University of Texas at El Paso}
  \streetaddress{500 W University Ave.}
  \city{El Paso} 
  \state{Texas} 
  \postcode{79968}}
\email{mhossain@utep.edu}

\renewcommand{\shortauthors}{R. Camacho Barranco \textit{et al.}}

\begin{abstract}
There is an overwhelming number of news articles published every day around the globe. Following the evolution of a news-story is a difficult task given that there is no such mechanism available to track back in time to study the diffusion of the relevant events in digital news feeds. The techniques developed so far to extract meaningful information from a massive corpus rely on similarity search, which results in a myopic loopback to the same topic without providing the needed insights to hypothesize the origin of a story that may be completely different than the news today. In this paper, we present an algorithm that mines historical data to detect the origin of an event, segments the timeline into disjoint groups of coherent news articles, and outlines the most important documents in a timeline with a soft probability to provide a better understanding of the evolution of a story. Qualitative and quantitative approaches to evaluate our framework demonstrate that our algorithm discovers statistically significant and meaningful stories in reasonable time. Additionally, a relevant case study on a set of news articles demonstrates that the generated output of the algorithm holds the promise to aid prediction of future entities in a story.


\end{abstract}

%
%



\maketitle
\section{Introduction}
The pervasiveness of the Internet has greatly facilitated editors to publish a large number of news articles everyday in any topic, practically without any restrictions of page-limits. Thus, the number of articles available to avid readers increases every day at a very fast pace. Many business and government organizations track these news articles to study public sentiments, business directions, progression of events, and many other aspects of economic and socio-political issues. However, tracking topics from different perspectives is a difficult task given that one news-story of today's interest could have evolved from another story of the past, thus forming a chain of relevant events. This leads to the concept of \textit{diffusion theory}~\cite{angulo_concepts_1980} which refers to the change of the distributional patterns of a phenomenon over time.
\begin{figure}[!b]
	\includegraphics[width=\columnwidth]{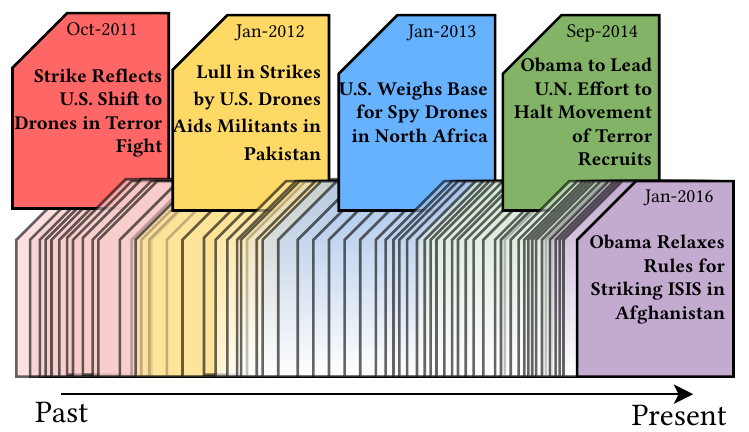}\\
	(a) A diffusion-based story.\\
	\includegraphics[width=\columnwidth]{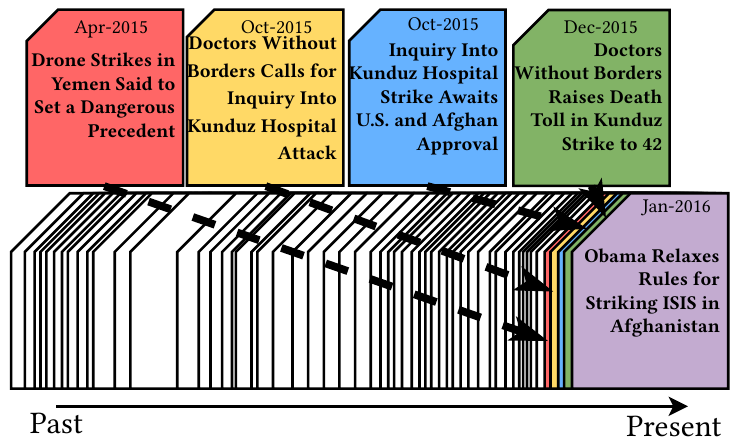}\\
	(b) A similarity-based story.
	\caption{ A diffusion-based approach captures documents from the past reflecting a smooth transition of the evolving story. In contrast, a similarity-based approach focuses on a narrow range of the timeline.
		\label{fig:diffusionandsimilarity}}
	\centering
\end{figure}

While capturing such diffusions over a timeline is still a challenge, researchers have targeted the problem of tracking stories in different guises, e.g., storytelling \cite{hossain_storytelling_2012,kumar_algorithms_2008}, storyboarding \cite{md_abdul_kader_connecting_2016}, connecting the dots \cite{shahaf_connecting_2010,hossain_connecting_2012}, and metro maps \cite{shahaf_information_2015}. Most of these methods find underlying connections between articles using a similarity-based network of documents. Their objective is mainly focused on generating a cohesive thread of a story. An extensive usage of cohesion may result in stories that revolve around a single \textit{chapter} of the evolving thread of the story. Figure~\ref{fig:diffusionandsimilarity} (a) illustrates a scenario generated using our diffusion-based approach, which captures the articles published in the past reflecting the evolution of the story. Figure~\ref{fig:diffusionandsimilarity}(b) presents the results obtained, for the same seed document, using a similarity-based approach to track past documents similar to the article of interest. While our approach brings out the underlying diffusion of concepts and their progression over time, the similarity-based approach provides a coherent story that revolves around a single event --- a drone strike in Kunduz, as shown in both examples. 



This paper presents a novel framework that takes as input a large corpus of documents, along with a user-specified set of seed documents. The framework runs through a palette of natural language processing (NLP) and other preprocessing tasks before it reaches its core objective function, which generates a storyline by finding and combining turning points that lead to the seed story. The end-product of the framework is a concise chain of documents from a multitude of news articles to help the user identify the evolution of a story, as shown in Fig.~\ref{fig:diffusionandsimilarity}(a).



The contributions of this paper can be summarized as follows:
\begin{itemize}
\item \textbf{Evolution:} We propose a novel method to discover a chain of documents published in the past given a set of seed documents, where the chain reflects evolution of the concepts in the seed documents.
\item \textbf{Diffusion and similarity:} Our proposed technique captures the underlying diffusion of concepts and has the flexibility to incorporate the similarity concepts from the state-of-the-art methods.
\item \textbf{Prediction:} We demonstrate the potential of the outcome generated by our approach as a tool to predict future states of an evolving story.
\item \textbf{Evaluation:} We conduct a set of experiments to quantitatively evaluate different aspects of our model. Additionally, a human participant-based study demonstrates that the discovered story lines were satisfactory.
\end{itemize}

\section{Related Work} \label{sec:related_work}
Several problems and tasks related to our work have been well studied. These problems include \textit{summarization}, \textit{event detection}, \textit{event extraction}, \textit{event evolution}, \textit{event prediction} and \textit{storytelling}, also known as \textit{connecting the dots}. \textit{Summarization} consists of extracting the most important sentences from a text corpus to help the user understand the main idea of the set of documents~\cite{allan_temporal_2001,radev_newsinessence:_2005,yan_evolutionary_2011,nenkova_survey_2012,binh_tran_structured_2013}. 

\textit{Event detection} consists of detecting temporal bursts of highly correlated documents~\cite{yang_learning_1999,yang_improving_2000,kleinberg_bursty_2002}, while \textit{event extraction} aims at obtaining useful knowledge regarding these events~\cite{kuzey_fresh_2014,rospocher_building_2016}. \textit{Event evolution}~\cite{yu_tracking_2015,wu_event_2016} and \textit{topic evolution}~\cite{jo_web_2011,kim_topic_2011,wang_understanding_2013} study how these abstract concepts (\textit{events} and \textit{topics}) evolve over time, based on latent relations within the corpus and temporal information encoded in each document.

The \textit{event prediction} through text mining problem has also been addressed in the literature. Radinsky \textit{et al.}~\cite{radinsky_learning_2012} described a system that uses \textit{causality extraction} to obtain pairs of terms that have a causal relation and that are later used to train a prediction algorithm. Luo \textit{et al.}~\cite{luo_measuring_2016} also address this problem by introducing the concept of \textit{semantic uncertainty}, which is used to estimate the \textit{most certain} next state based on the current state, or event. This approach is particularly useful when there is limited historical data available.

The main component of our framework is focused on the problem of \textit{storytelling}, or \textit{connecting the dots}. There are several approaches that aim to solve this problem for different types of datasets, such as scientific articles~\cite{hossain_connecting_2012}, entity networks~\cite{faloutsos_fast_2004,fang_rex:_2011,hossain_helping_2011}, image collections~\cite{heath_image_2010,suen_nifty:_2013} and document collections~\cite{nallapati_event_2004,kumar_algorithms_2008,shahaf_connecting_2010,ahmed_unified_2011,gillenwater_discovering_2012,zhu_finding_2014,shahaf_information_2015,gu_personalized_2016,wu_event_2016}. Most of the work on this problem uses graph-based representations of a document or entity set~\cite{faloutsos_fast_2004,nallapati_event_2004,hossain_storytelling_2012,gillenwater_discovering_2012,zhu_finding_2014}. \textit{Storytelling} has several different applications, such as intelligence analysis~\cite{hossain_helping_2011,hossain_storytelling_2012}, news recommendation~\cite{gu_personalized_2016}, search engines~\cite{fang_rex:_2011,zhu_finding_2014}, and social network analysis~\cite{faloutsos_fast_2004,suen_nifty:_2013,ning_uncovering_2015}.

Solutions to the \textit{connecting the dots} problem include the use of probabilistic approaches, such as random walks~\cite{shahaf_connecting_2010,zhu_finding_2014,shahaf_information_2015,gu_personalized_2016,wu_event_2016}, Monte Carlo simulations~\cite{ahmed_unified_2011} and determinantal point processes (DPP)~\cite{gillenwater_discovering_2012}. Shahaf \textit{et al.}~\cite{shahaf_connecting_2010} introduced the concept of coherence and coverage to assess the quality of a chain limited by two user-specified endpoints, and then extended this work by building \textit{metro maps}~\cite{shahaf_information_2015} which are formed by several coherent chains that intersect at some points, forming a map-like structure. Gillenwater \textit{et al.} also obtain several coherent chains using a DPP-based model.

Our work has several differences with respect to the previously mentioned methods. First, we use an optimization-based approach similar to the one presented by Shahaf \textit{et al.}~\cite{shahaf_connecting_2010}, but we introduce the concept of \textit{diffusion} to identify the evolution of the concepts in the seed documents. The use of diffusion allows our framework to create stories based on a combination of different events, which is not possible using previously proposed methods.
Second, our work discovers the most relevant documents associated with each turning point to provide a concise summary for each turning point event. Third, we leverage the topic distribution of each document to reduce our search space while keeping important documents for our optimization.
Finally, we suggest the possibility of using the resulting relevant documents to build a regression framework for entity prediction.
\begin{figure}
	\includegraphics[width=\linewidth]{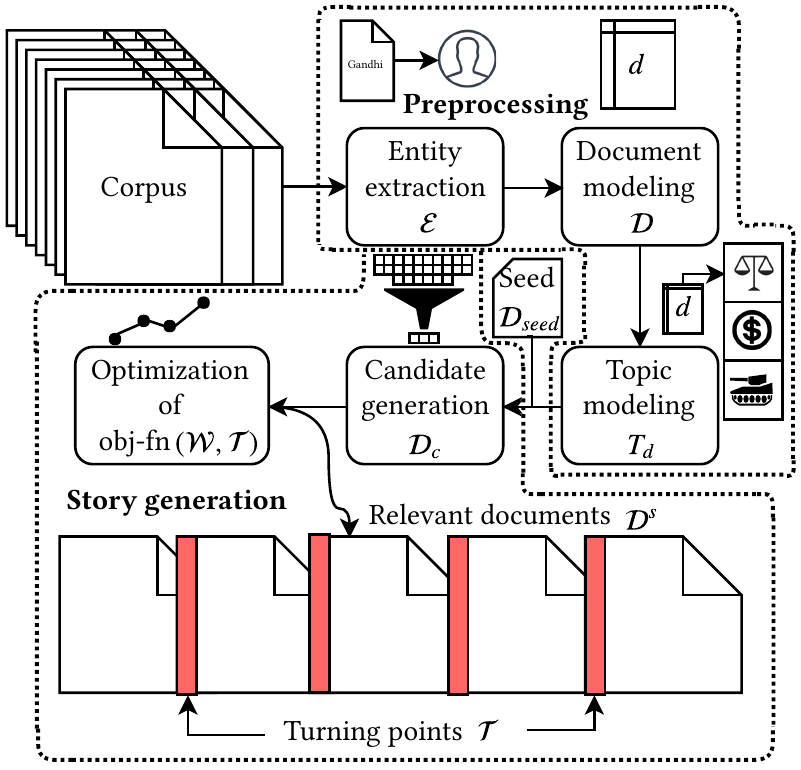}
	\caption{Proposed framework. \label{fig:framework}}
	\centering
\end{figure}

\section{Problem Description} \label{sec:problem}
This work focuses on news articles and the entities within the news. The entities detected are persons, organizations, and locations. 
Let $\mathcal{D} = \{d_1, d_2,\ldots, d_{|\mathcal{D}|}\}$ be the set of documents and $\mathcal{E} = \{e_1, e_2,\ldots, e_{\mathcal{E}}\}$ be the set of entities in the news corpus. Each document $ d \in \mathcal{D} $ contains a set of entities $ \mathcal{E}_d \subset \mathcal{E}$. Additionally, each document $d$ contains its publication date $t_d$. As part of the preprocessing steps, we apply topic modeling using Latent Dirichlet Allocation (LDA)~\cite{blei_latent_2003} on $\mathcal{D}$, obtaining a topic distribution $T_d$ for each document $d$.

\subsection{What is the expected outcome?} \label{sec:problem:outcome}
Let the user input be a set of seed documents $\mathcal{D}_{seed} \subset \mathcal{D}$, where $|\mathcal{D}_{seed}| \geq 1$. We define a \textit{turning point} in the story $\tau \in \mathcal{T}$ as a specific date in which the story under analysis has a significant change. Let $\mathcal{S}$ be a vector where each element, defined as \textit{segment}, is a pair of consecutive turning points such that $|\mathcal{S}| = |\mathcal{T}| - 1$, and $\mathcal{S} = [(\tau_1,\tau_2), (\tau_2, \tau_3),\cdots,(\tau_{|\mathcal{T}|-1}, \tau_{|\mathcal{T}|})]$.  

Our main goal is to split the set of documents in $|\mathcal{S}|$ segments by finding $|\mathcal{T}|-2$ turning points, since the first and last turning points are fixed. For each segment $s \in \mathcal{S}$ we want to find a subset of documents $\mathcal{D}^{s} \subset \mathcal{D}$ that help the user study the evolution of the prominent entities in $\mathcal{D}_{seed}$ through various different events. 

\section{Methodology} \label{sec:method}

Our framework comprises two main stages: (1) preprocessing, which creates document and topic models from a text corpus and (2) story generation, which takes a set of seed documents and other constraints as inputs and, via an optimization routine, outputs a story formed by relevant documents that have a common thread, but that belong to different events. Fig.~\ref{fig:framework} illustrates this framework.
\subsection{Preprocessing}
In the preprocessing stage, our framework: (1) extracts entities (e.g., person, location, organization) from the documents, (2) represents each document as a vector of entities, and (3) obtains a topic distribution for each document in our corpus.
We extract entities from the text corpus using standard Named Entity Recognizers~\cite{alias-i_lingpipe_2008,finkel_incorporating_2005}. 
Our framework leverages a tf-idf model with cosine normalization \cite{hossain_storytelling_2012} to generate weights $w(e, d)$ for each entity $e \in \mathcal{E}$ in each document $d \in \mathcal{D}$. 


\subsection{Story generation} \label{sec:method:candidate}
In this stage, our framework furnishes a candidate set of documents $\mathcal{D}_c \subset \mathcal{D}$ from the entire dataset that satisfy some temporal and topical constraints with respect to the seed documents $\mathcal{D}_{seed}$. The temporal and topical constraints are driven by user input regarding how far back in time the algorithm should track to detect an origin, or, how much deviation the algorithm should allow in terms of topic distribution as compared with the seed set. 

The temporal criteria establishes that all of the candidate documents must have been published before the most recent seed document and within a certain minimum threshold $t_{\min}$, i.e., $d_i$ can only be part of $\mathcal{D}_c$ if $t_{\min}< t_{i} < \max(t_{seed})$, where $t_i$ is the publication date of document $d_i$.
The topical criteria expresses, in terms of topic distributions, how much deviation the candidate documents can have from the seed documents when optimizing for a diffusion and cohesion objective. For each document $d_{seed} \in \mathcal{D}_{seed}$, we compute the KL-divergence~\cite{kullback_information_1951}, or \textit{topical divergence}, between the topic distribution of the seed document $T_{d_{seed}}$ and the $T_d$ of all the documents $d \in \mathcal{D}$. Document $d$ is included in $\mathcal{D}_c$, only if $\forall d_{seed} \in \mathcal{D}_{seed} \left( \text{KL-divergence}(T_d,T_{d_{seed}}) \leq \alpha \right)$, where $T_d$ is the topic distribution of document $D$, obtained using the LDA \cite{blei_latent_2003} and $\alpha$ is a user-defined parameter. In Section~\ref{sec:method:formalizing}, we will refer to $\mathcal{D}_c$ by $\mathcal{D}$, for simplicity.

Finally, the framework optimizes fitment of a story model that identifies turning points of the story over a timeline and provides relevant documents within each segment. The outcome of the optimization routine helps understand the evolution of the news-story described in the seed document(s). This essential is further described in Section~\ref{sec:method:formalizing}.



\section{Formalizing Our Approach} \label{sec:method:formalizing}
In this section, we gradually introduce the concepts and formulations that give shape to our model, which has the goal of obtaining a coherent story from separate events. We present our ideas iteratively and explain why and how new penalty terms were developed to achieve our goal.

\subsection{Defining incoherence}
The objective is to ensure that the documents within each time segment of $\mathcal{S}$ are coherent. A simplistic approach is to minimize the distance of all pairs of documents within each time segment. For simplicity, we use the notation $d^s$, to indicate the segment $s$ to which document $d$ belongs to, i.e., $d^{1}$ is defined as a document in the segment between $\tau_1$ and $\tau_2$. We define this distance as incoherence:

\begin{equation} \label{eq:incoherence1}
\text{incoherence}_1(s) = \frac{\sum_{i \neq j}^{|P^s|}{||d_i^s-d_j^s||_{\varrho}}}{|P^s|}
\end{equation}
\noindent where $\mathcal{S}$ is the set of segments, $d^s$ is a document s.t. $d^s \in D^s$, where $D^s$ is the set of candidate documents that are within segment $s$. $P^s$ is a set of pairs $(i, j)$ that represent the different 2-permutations of the indices of pairs of documents $d_i^s$ and $d_j^s$, computed using the following equation.



\begin{equation} \label{eq:perms}
|P^s|=\frac{|D^s|!}{2(|D^s|-2)!}=\frac{|D^s|^2-|D^s|}{2}
\end{equation}

\subsection{Defining a distance metric and forming an initial objective function} \label{subsec:v1}
In Eq.~\ref{eq:incoherence1}, the distance computation between the weight vectors of two documents is represented as $||\cdot||_\varrho$, where $\varrho$ is substituted by the distance measure chosen. We prefer the Soergel distance between a pair of documents $d_i$ and $d_j$, defined in Eq.~\ref{eq:soergel}.

\begin{equation} \label{eq:soergel}
\text{soergel}(d_i, d_j) = \frac{\sum_{e \in \mathcal{E}}\left|w(d_i,t)-w(d_j,t)\right|}{\sum_{e \in \mathcal{E}}\max\left(w(d_i,t),w(d_j,t)\right)}
\end{equation}
\noindent where $\mathcal{E}$ is the set of all the entities that appear in the dataset, and $w(d, e)$ is the tf-idf weight of entity $e$ for document $d$. 


The initial objective function is thus defined as Eq.~\ref{eq:v1}, where $s$ represents the current segment.
\begin{equation} \label{eq:v1}
\mathcal{F}_1(\mathcal{T}) = \sum_{s=1}^{|\mathcal{S}|}{\text{incoherence}}_1(s)
\end{equation}

This initial objective function is not enough to serve the purpose of capturing evolution in terms of diffusion. Eq. \ref{eq:v1} tends to generate one very large segment and many small ones, which is undesirable. Ideally, two documents that were published at distant dates should not belong to the same segment. 
Another limitation of the approach in Eq. \ref{eq:v1} is that it uses hard assignments of documents to a single segment, while we would prefer a soft membership of documents to time segments for better analytic flexibility. 


\subsection{Including penalty based on time segmentation}
To tackle the problem of generating small segments, we include a new term in our definition of \textit{incoherence}. The new term helps by avoiding segments containing document pairs with distant publication dates. Therefore, a solution with many small segments or a solution with a very large segment will not be the optimal. This term is simply a date difference penalty ($\text{date}_\Delta$). 
\begin{equation} \label{eq:date_diff}
\text{date}_\Delta(i, j)=|t_{i}-t_{j}|
\end{equation}
\noindent where $t_{i}$ is the publication date of document $d_{i}$. Although Eq. \ref{eq:date_diff} provides date differences, in practice, we normalize the values to a smaller range and the user can flexibly modify the normalization range.
With this new term, we can redefine incoherence as Eq.~\ref{eq:incoherence2}. 
\begin{equation} \label{eq:incoherence2}
\text{incoherence}_2(s) = \frac{\sum_{i < j}^{|P^s|}{\text{soergel}(d_i, d_j)*\text{date}_\Delta(i, j)}}{|P^s|}
\end{equation}
The objective function has the same form of Eq.~\ref{eq:v1}, but uses the new incoherence equation. The advantage of this new formulation is that it reduces the incidence of very small segments significantly. However, the current objective function is not capable of capturing the essence of \textit{diffusion} or \textit{turning point} because it only takes into account coherence of each segment. This can result in very similar segments that do not show the evolution of a story. Another disadvantage is that a document can only be assigned to a single segment, and thus only discrete optimization routines can be used.

\subsection{Introducing unconnectedness}
While we prefer similar documents within a segment, we expect high dissimilarity between the documents of different segments. We call this property of diffusion across segments \textit{unconnectedness}. We expect high \textit{unconnectedness} as expressed by Eq.~\ref{eq:unconn}. 

\begin{equation}\label{eq:unconn}
\text{unconnectedness}(s) =\frac{\sum_{i=1}^{|D^s|}\sum_{j=1}^{|G^s|}{\text{soergel}(d_i^s, g_j^s)}}{|D^s|*|G^s|}
\end{equation}
\noindent where $G^s = D - D^s$ is the set of documents that are not within segment $s$, and $g^s \in G^s$.
Combining $\text{unconnectedness}$ with the original objective function, we obtain the following equation.
\begin{equation} \label{eq:v2_distance}
\mathcal{F}_2(\mathcal{T})=\sum_{s=1}^{|\mathcal{S}|}{\left(\text{incoherence}_2(s)-\text{unconnectedness}(s)\right)}
\end{equation}
In practice, we normalized and rescaled these two terms to adjust the importance of minimizing incoherence and maximizing unconnectedness. However, a major problem with this formulation is that because of the conflicting objectives, we cannot easily define which penalty should have the largest effect. 

\subsection{Changing to segment similarity} \label{subsec:v3}
To eliminate the conflicting objectives issue, and to obtain a smooth formulation, we introduce the concept of similarity between segments. This penalty, formalized as Eq.~\ref{eq:similarity1}, will measure how similar documents from one segment are with respect to the documents in the other segments.

\begin{equation} \label{eq:similarity1}
\text{similarity}_1(s) = \frac{\sum_{i=1}^{|D^s|}\sum_{j=1}^{|G^s|}{\mathrm{e}^{-{\text{soergel}(d_i^s, g_j^s)}}}}{|D^s|*|G^s|}
\end{equation}

We seek to minimize $\text{similarity}(s)$ for each segment $s$. The updated objective function is:

\begin{equation} \label{eq:v2}
\mathcal{F}_3(\mathcal{T})=\sum_{s=1}^{|\mathcal{S}|}{\left(\text{incoherence}_2(s) * \text{similarity}_1(s)\right)}
\end{equation}
Eq. \ref{eq:v2} allows minimization of incoherence and similarity simultaneously without keeping track of the importance of each other since they are represented as factors. However, this formulation still suffers from two issues: (1) a news article can only belong to a single segment, which means only discrete optimization routines can be used, and (2) it is possible to have several articles that are irrelevant to the story within each segment. We address these issues in the following subsections.
\subsection{From discrete to continuous}
The objective function of Eq. \ref{eq:v2} considers  discrete assignments of documents to time segments. This results in lesser flexibility in terms of smoothness and use of optimization routines. To indicate the certainty that a document with time stamp $t$ belongs to a continuous range of time, in our case a segment, where $t^s_L$ is the lower limit or a turning point and $t^s_H$ is the upper limit, a membership function can be defined as:

\begin{equation} \label{eq:gamma}
\gamma(t, t^s_L, t^s_H)=
     \begin{cases}
       \frac{1}{\sqrt[]{2\pi \hat{\sigma}^2}}\mathrm{e}^{\left(-\frac{(t - t^{s}_L -\hat{\sigma}^2)^2}{2 \hat{\sigma}^2}\right)} &\quad\text{if } t\le t^s_L\\
       \frac{1}{\sqrt[]{2\pi\hat{\sigma}^2}}
 &\quad\text{if } t^s_L < t < t^s_H\\
       \frac{1}{\sqrt[]{2\pi \hat{\sigma}^2}}\mathrm{e}^{\left(-\frac{(t - t^{s}_H + \hat{\sigma}^2)^2}{2 \hat{\sigma}^2}\right)} &\quad\text{if } t^s_H\le t
     \end{cases}
\end{equation}

\begin{figure}
	\includegraphics[height=1in]{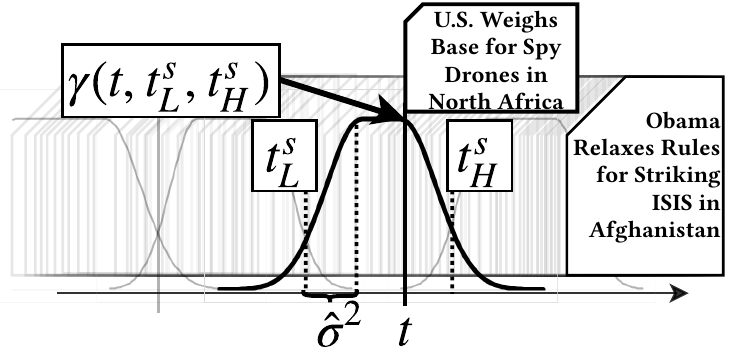}
	\caption{The $\gamma(t, t^s_L, t^s_H)$ function as described in Eq.\ref{eq:gamma}. \label{fig:gamma}}
	\centering
\end{figure}

Although Eq.\ref{eq:gamma} is partitioned, it is continuous and smooth because the membership function is formed by two halves of different Gaussian distributions, and a linear section that connects both, as shown in Fig.~\ref{fig:gamma}. The means of the distributions are the lower and higher limits, and the standard deviation $\hat{\sigma}$ for both is selected by the user to modify the shape of the membership function. This score is not a probability, but we can obtain the probability that time stamp $t$ belongs to segment $s$ by dividing the score over the sum of the membership values of $t$ for all segments in $\mathcal{S}$.

With this new concept of membership function, we must update our previous definitions. 
\begin{equation}
\text{incoherence}_3(s) = \frac{\sum_{i, j}^{|D| \times |D|}{\Phi *\text{soergel}(d_i, d_j)*\text{date}_\Delta(i, j)}}{\sum_{i, j}^{|D| \times |D|}{\Phi}}
\end{equation}
where
\begin{equation}
\Phi = \gamma(t_{i}, t^s_L, t^s_H)*\gamma(t_{j}, t^s_L, t^s_H)
\end{equation}
Segment similarity can be redefined as:
\begin{equation}
\text{similarity}_2(s) = \frac{\sum_{i,j}^{|D|\times |D|}{\phi*\mathrm{e}^{-{\text{soergel}(d_i, d_j)}}}}{\sum_{i,j}^{|D|\times |D|}{\phi}}
\end{equation}
where
\begin{equation}
\phi = \gamma(t_{i}, t^s_L, t^s_H)*\left(1-\gamma(t_{j}, t^s_L, t^s_H)\right)
\end{equation}

The objective function has the same form of Eq.~\ref{eq:v2}, but uses the new incoherence and similarity equations.
The main advantage of this new formulation is that it is suitable to be used with a larger variety of optimization routines. However, a new problem is introduced with this objective function, which is that it tends to overlap the turning points. Furthermore, we still have the problem of many documents being irrelevant to a segment.

\subsection{Introducing segment overlap penalty}
\begin{figure}
	\includegraphics[]{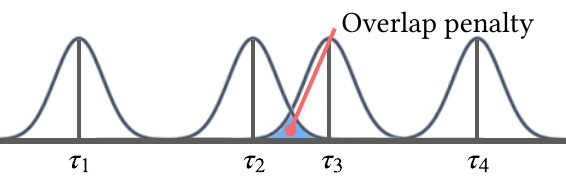}
	\caption{Potential overlap of turning points. \label{fig:overlap}}
	\centering
\end{figure}

To avoid overlap of segments (as shown in Fig.~\ref{fig:overlap}) we include a penalty that is high when two turning points are very close to each other. It is defined as:
\begin{equation} \label{eq:overlap}
\text{overlap}=\left(1 + \sum_{m,n, m < n}^{|\mathcal{S}|-1\times |\mathcal{S}|-1}{e^{\left(-\frac{(t^{m} - t^{n})^2}{2 \sigma^2}\right)}}\right)
\end{equation}
\noindent where $t^m$ and $t^n$ are the publication date of two different documents, and $\sigma$ is a hyperparameter that allows the user to set a threshold defining the degree of the allowed overlap. 

The objective function with the $\text{overlap}$ penalty is: 
\begin{equation} \label{eq:v3}
\mathcal{F}_4(\mathcal{T}) = \sum_{t=1}^{|\mathcal{S}|}\left(\text{incoherence}_3(s)*\text{similarity}_2(s)\right)*\text{overlap}
\end{equation}

\subsection{Adding weights as a relevance factor}

It is difficult for a user to analyze a massive amount of candidate documents given only the time segments. Our framework mitigates this problem by providing a score with each document representing how important it is for this study. This score is a variable for the optimization routine and it is used in the objective function. The variable is represented as a vector $\mathcal{W}$ of length equal to the number of candidate documents. A weight of 1.0 indicates that the document is relevant and 0.0 indicates that it is not. 

The updated definitions are Eq.~\ref{eq:incoherence3} and Eq.~\ref{eq:similarity2}, where $w_i \in \mathcal{W}$ is the weight for document $d_i$. 

\begin{equation}\label{eq:incoherence3}
\text{incoherence}_4(s) = \frac{\sum_{i, j}^{|D| \times |D|}{w_i * w_j * \Phi *\text{soergel}(d_i, d_j)*\text{date}_\Delta(i, j)}}{\sum_{i, j}^{|D| \times |D|}{w_i * w_j * \Phi}}
\end{equation}

\begin{equation} \label{eq:similarity2}
\text{similarity}_3(s) = \frac{\sum_{i,j}^{|D|\times |D|}{w_i * w_j * \phi * \mathrm{e}^{-\text{soergel}(d_i, d_j)}}}{\sum_{i,j}^{|D|\times |D|}{w_i * w_j * \phi}}
\end{equation}
The objective function has the same form of Eq.~\ref{eq:v3}, but uses the new incoherence and similarity formulations and has $\mathcal{W}$ as input. However, a new problem emerges --- the optimal value is when all the weights are set to zero.

\subsection{Uniformity penalty}
To ensure that only a subset of $\mathcal{D}$ is truly relevant, and to avoid the degenerate case where the optimal value is with all the weights set to zero, we need to minimize the uniformity of the weights vector $\mathcal{W}$. A completely uniform vector $\mathcal{W}$, i.e., all zeros or all ones, will result in a high penalty, while a completely non-uniform weights vector will result in no penalty. The uniformity penalty is: 
\begin{equation} \label{eq:uniformity}
\text{uniformity} = \left(1 + \sum_{s=1}^{|\mathcal{S}|}{\left(1 - \frac{\left(\left|\left|\frac{\mathcal{W}_s * \Gamma_s^\top}{\sum{\mathcal{W}_s * \Gamma_s^\top}}\right|\right|_2*\sqrt{|\mathcal{W}_s|}\right)-1}{\sqrt{|\mathcal{W}_s|}-1}\right)}\right)
\end{equation}
\noindent where $\Gamma_e$ is a vector of values returned by the membership function (Eq.~\ref{eq:gamma} for the documents that fall in segment $s$).

\subsection{The final objective function} \label{sec:method:optimization}
The final objective function is shown below.
\begin{equation} \label{eq:v4}
\mathcal{F}_5(\mathcal{T}, \mathcal{W}) = \sum_{s=1}^{|\mathcal{S}|}\left(\text{incoherence}_4(s)*\text{similarity}_3(s)\right)*\text{overlap}*\text{uniformity}
\end{equation}


The objective function is minimized for two vectors: $\mathcal{S}$ and $\mathcal{W}$. The elements $s \in \mathcal{S}$ are bounded between $[0, date_{max}]$ where $date_{max}$ is the value representing the publication date of the most recent news article while 0 represents the earliest document on the candidate dataset. The elements $w \in \mathcal{W}$ are bounded between $[0, 1]$, as stated earlier. We used the quasi-newton limited memory algorithm for bound constrained optimization (L-BFGS-B)~\cite{zhu_algorithm_1997} to minimize our objective function. 
\section{Experimental Results}
For our experiments, we use a corpus of 400,842 New York Times articles published between January 2000 and June 2016. The corpus contains 3,320,886 unique entities. To the best of our knowledge, there are no publicly available labeled benchmark datasets that can allow us to perform a supervised evaluation of our framework. Our evaluations are based on a combination of statistics- and human participant-based studies. We seek to answer the following questions.

\begin{enumerate}
\item How well does our approach reflect the evolution of a story? (Section~\ref{sec:exp:contin})
\item How does our method compare to other methods such as clustering and similarity-based storytelling? (Section~\ref{sec:exp:comp})
\item How do the various proposed regularizations impact story quality? (Section~\ref{sec:use})
\item How does the statistical significance for the position of the turning points change while modifying the different hyper-parameters of our optimization? (Section~\ref{sec:exp:sig})
\item What is the repeatability of the optimization? (Section~\ref{sec:exp:rep})
\item Can we use the set of highly relevant documents with respect to a seed document to predict which entities are expected to appear in a future document? (Section~\ref{sec:exp:pred})

\end{enumerate}


\subsection{Evaluation of Chain Continuity} \label{sec:exp:contin}
The lack of a benchmark dataset makes it very difficult to evaluate the performance of our optimization framework. To address this issue, we designed a user study motivated by the work of Shahaf \textit{et al.}~\cite{shahaf_connecting_2010}. We picked four different topics and two different random documents related to each topic as seed documents. The topics selected were \textit{aviation}, \textit{Brexit}, \textit{ISIS}, and \textit{tensions between North and South Korea}. Four evaluators were asked to score three topics each, so that each topic had a total of six evaluations\footnote{Evaluation questions available at: \url{https://storyeval.herokuapp.com/}}.

For each segment, we selected the document with the highest weight. Four past documents were selected for each seed document. Users evaluated four criteria on a scale of 1 to 5, with five being the best, for each sequence: 
\begin{itemize}
\item \textit{Familiarity:} How familiar the person is with this topic.
\item \textit{Relevance:} Do the documents seem relevant to the topic?
\item \textit{Coherence} Is the chain coherent? A lower score is provided if there is a document in the chain that does not belong to the topic.
\item \textit{Broadness:} Does each document provide new information guaranteeing a slight shift in topic over time?
\end{itemize}

\begin{figure}
\includegraphics[width=\linewidth]{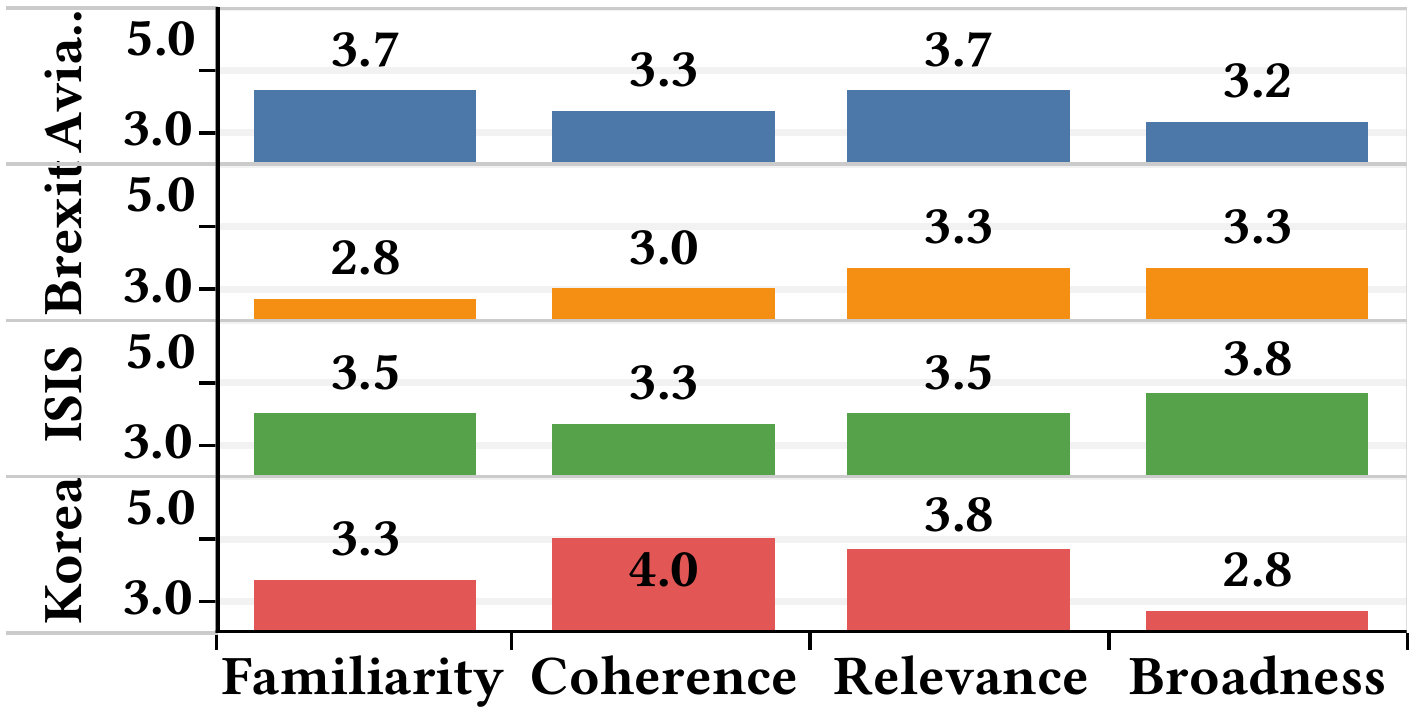}
\caption{ Average scores obtained for each metric per topic, a higher score is better (max. 5.0).  \label{fig:human_range}}
\centering
\end{figure}

The average scores per criterion are presented in Fig.~\ref{fig:human_range}. The results indicate that our algorithm performs, on average, above a score of 3, which indicates that the chains are coherent, relevant, and broad enough to capture evolution. An interesting observation is that the \textit{Brexit} topic, which had the lowest average familiarity score, also has the lowest scores for coherence and relevance, indicating that such a study might be biased toward the familiarity of the topic.

\subsection{Comparison with Other Methods} \label{sec:exp:comp}
We compare the outcomes of our framework with two other methods: (1) a $k$-means clustering, which combines the time feature with entities and (2) a similarity-based approach where consecutive nearest neighbors are selected as a chain starting with the seed document. We selected one best document per time-segment for this evaluation using our method. For the $k$-means clustering based approach we selected one document closest to the centroid of each cluster as a relevant document. For the similarity-based approach, we used the sequence of documents generated from the seed document as a path. Notice that the two alternative approaches focus on distance or similarity, while our objective function is designed to capture diffusion. Hossain \textit{et al.} introduced the concept of \textit{dispersion coefficient} in evaluating the Storytelling algorithm \cite{hossain_storytelling_2012}. We use the same concept to evaluate the quality of a chain of documents $\{d_0$, $d_1$, . . ., $d_{n-1}\}$, containing $n$ articles. 

\begin{equation*}
	\psi = 1 - \frac{1}{n-2}\sum_{i=0}^{n-3}\sum_{j=i+2}^{n-1} \text{disp}(d_i, d_j)
\end{equation*}
\noindent where
\begin{equation*}
	\text{disp}(d_i, d_j) =
	\begin{cases}
		\frac{1}{n+i-j} &\quad\text{if soergel}(d_i, d_j) < \theta \\
		0 &\quad\text{otherwise}
	\end{cases}
\end{equation*}

The dispersion coefficient of a chain of documents is 1.0 only if consecutive pairs of documents meet a distance threshold $\theta$. The coefficient is 0 when every pair of documents in the chain satisfies the distance threshold.

\begin{figure}
\includegraphics[width=\columnwidth]{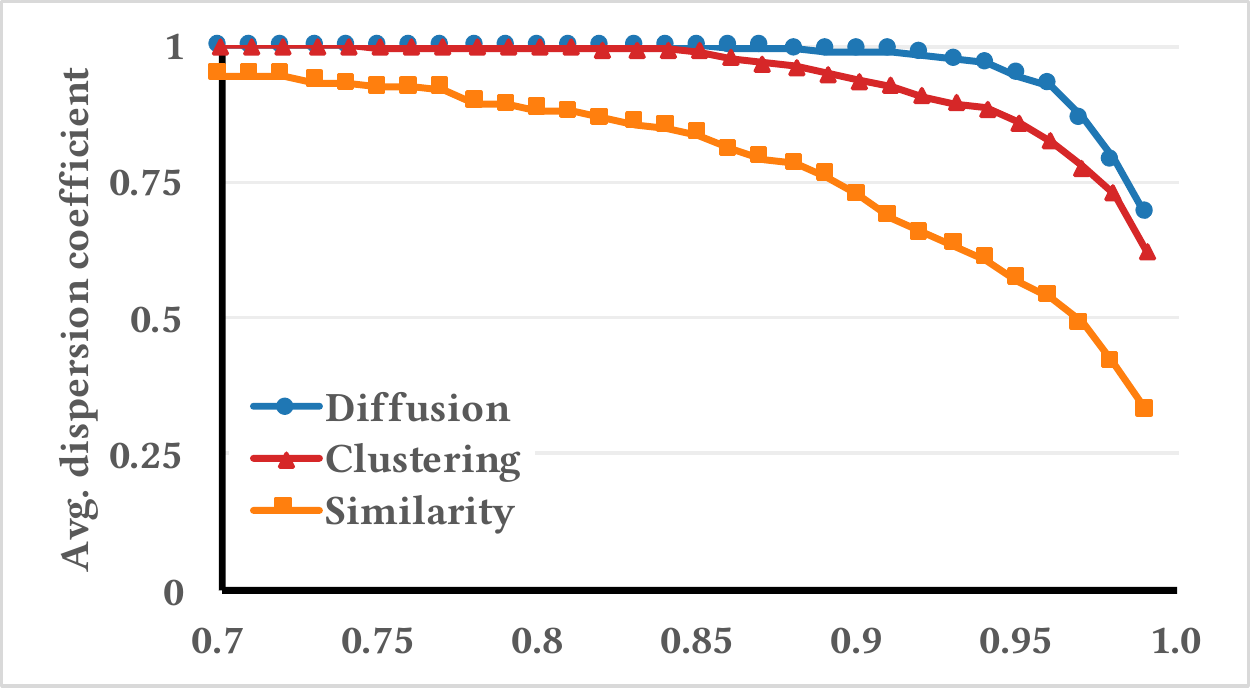}
\caption{ Comparison of average dispersion coefficient for stories created using our diffusion based approach, a clustering based algorithm, and a similarity based technique. Our approach results in the best (highest) dispersion. \label{fig:diffusion}}
\centering
\end{figure}

We generated chains for 116 different seed documents using all three approaches under consideration. Figure~\ref{fig:diffusion} compares the average dispersion coefficient versus the distance threshold. Figure~\ref{fig:diffusion} shows that the average dispersion of our diffusion-based method provides the highest dispersion coefficients for any distance threshold. This indicates that our method generates chains with smooth transition of topics, which is our goal.

\subsection{Use Case Study} \label{sec:use}
\begin{figure}
\includegraphics[width=\linewidth]{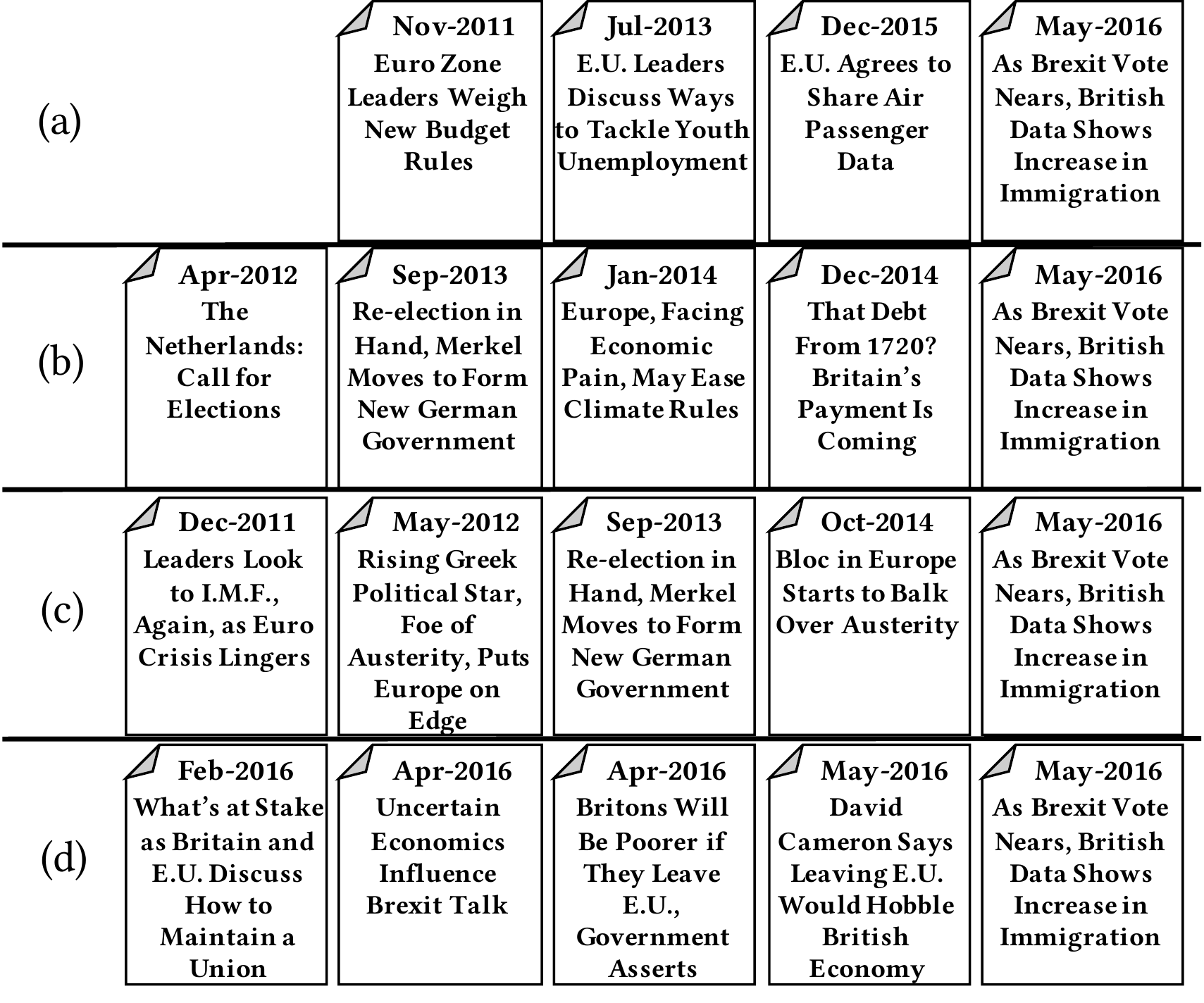}
\caption{Stories generated using (a) the objective function presented in Sec.~\ref{subsec:v1}, (b) the formulation presented in Sec.~\ref{subsec:v3}, (c) our diffusion-based approach and (d) a similarity-based approach.   \label{fig:stories}}
\centering
\end{figure}

In this experiment, we perform a use case study to compare the results of our diffusion-based approach with two preliminary versions of our objective function, as well as with the similarity-based approach described in Sec.~\ref{sec:exp:comp}. We execute each of these methods for the same seed document and select the document in the middle of each segment. The number of segments is set to 5. 
Fig.~\ref{fig:stories}(a) shows the story generated using the formulation presented in Sec.~\ref{subsec:v1}. In this case, because the date penalty was not included, one event overlaps with the other, and thus only four segments are obtained. The topics of each document are about the European Union, but lack a common thread. Fig.~\ref{fig:stories}(b) shows a story generated using by optimizing the objective function presented in Sec.~\ref{subsec:v3}, which is in the continuous space and includes the date and similarity penalties. An analysis of this story shows that there is no coherent thread, which is caused by the lack of weights that indicate the relevance of a document.

Fig.~\ref{fig:stories}(c) shows the story resulting from our diffusion-based approach. At first glance, some of the articles seem disconnected from a continuous thread. However, after a brief analysis, we can find a very relevant thread: the first article is about the financial assistance required by Greece and Portugal because of the severe crisis. The second article mentions how Germany made efforts to keep Greece in the Euro zone, because of fear of the consequences of Greece leaving. The next article mentions the future plans of the German chancellor with respect to the European Union and the imposed austerity, and in particular mentions that Britain's prime minister expected the support from Ms. Merkel. The article before the seed document mentions the latest European crisis, and how several countries were ``rebelling'' against the German-mandated austerity. The seed document talks about immigration being one of the main motivations for Brexit. The article mentions that there was an influx of immigrants on 2015, presumably escaping from the economical crisis in their countries of origin. This is a coherent and relevant thread consisting of different events, which was our ultimate goal.

The story obtained using the similarity-based approach, shown in Fig.~\ref{fig:stories}(d), is also coherent and relevant. However, the story completely revolves around Brexit, and a simple keyword search about the topic would probably return similar results. Thus, we consider that our result provides a better insight into the possible paths that lead to the event described in the seed document.

\subsection{Statistical Significance Analysis} \label{sec:exp:sig}
Evaluation using precision, recall, and other variants does not apply in the context of our work because of the lack of labeled datasets suitable for our task.
An alternative approach, other than the user studies and quality-evaluation using an unsupervised metric like dispersion coefficient, is to use a statistical analysis of the significance (p-value) as a sanity check to make sure that our formulation obtains results that have a very low probability of being obtained by random chance.
Our objective function has several user-modifiable parameters that can change analytic viewpoints. In this section, we will explore how the significance changes with varying parameters.
 


For our problem, we defined $p_{TP}$ as the p-value for the turning points vector $\mathcal{T}$. We generated $m$ random samples $\mathcal{T}_S$ which are of the same size as $\mathcal{T}$. Then, we compared each of the $m$ random samples with $\mathcal{T}$ by computing their element-wise difference and comparing it to a tolerance $\beta$, i.e., $(\mathcal{T}_s-\mathcal{T}) \leq \beta$. We counted the number of times this condition was satisfied in an auxiliary variable $a$. Finally, we returned $p_{TP} = \frac{a}{m}$. 

The experiments were conducted by computing the $p_{TP}$ of several different configurations. We precomputed 100,000 random samples. Then, we used a number of random seed document sets and ran the optimization routine for different configurations. We finally averaged the $p$ values across the documents.



\begin{figure*}

\includegraphics[width=\linewidth]{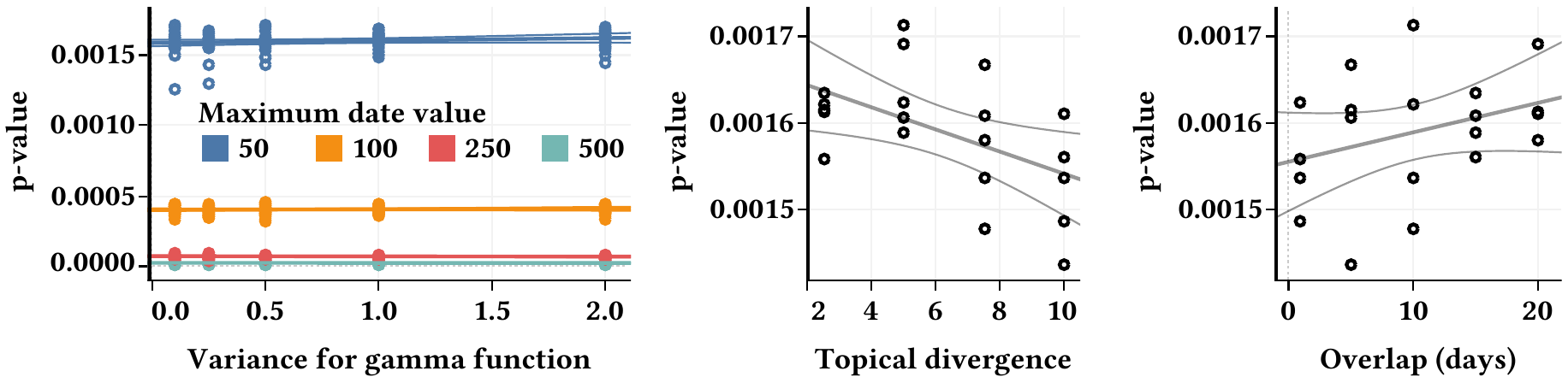}
\begin{flushleft}
\hspace{3.8cm}\textbf{(a)}\hspace{5.9cm}\textbf{(b)}\hspace{5.1cm}\textbf{(c)}
\end{flushleft}
\caption{Significance of turning points vector vs. (a) variance of gamma function, (b) topical divergence and (c) date overlap penalty. Each data series includes a predicted trend line as well as upper and lower 95\% confidence lines. \label{fig:significance}}
\centering
\end{figure*}

One of the configurations relates to the publication dates of the articles. The dates are scaled to a continuous range. In Fig.~\ref{fig:significance}(a), it is evident that the $p_{TP}$ value significantly decreases while increasing the \textit{maximum date} value. This is because the range for the turning points increases with the scaling. Fig.~\ref{fig:significance}(a) also shows that our approach generates statistically significant results with any value of the variance for the gamma function $\hat{\sigma}^2$, and that the change in significance while varying this parameter is limited.

The standard deviation parameter for overlap $\sigma$ controls the width of the bell-shaped curve shown in Fig.~\ref{fig:overlap}. A penalty is added if any of the previous turning points fall within this curve. Figure~\ref{fig:significance}(c) shows that as the overlap $\sigma$ increases, the $p_{TP}$ value increases. The maximum topical divergence parameter is used during the candidate generation stage to filter documents that are above this value. Figure~\ref{fig:significance}(b) shows that when the topical divergence increases, the $p_{TP}$ value decreases, albeit slightly. In general, our approach generates statistically significant results even with varying parameters.





\subsection{Local Optimization: Repeatability of Concepts } \label{sec:exp:rep}
One of the crucial elements of local optimization, as used in our framework, is that it may produce multiple results with different executions for the same set of seed documents and configurations. From an analytic perspective, multiple results for the same seed set provide a deeper understanding of the evolution. However, too much diversity in the results for the same seed set may be overwhelming for users. In this experiment, we analyze the repeatability of the results, in particular for the turning points, for the same set of seed sets.  We selected random seed documents from different topics and repeated the optimization 100 times for each seed keeping track of the resulting turning-point vector, each time with a different initialization for the optimization routine. The collected vectors were compared pairwise and grouped into \textit{buckets} based on similarity. To define similarity in this context, we first define a match as when the distance between two turning points from different vectors is below a threshold, $\zeta$. Two vectors are similar if the number of matches is above a \textit{minimum match} parameter. Ideally, the number of buckets should be small (close to 1). If the number of buckets is 100, this means that none of the pairs of vectors fulfill the conditions mentioned above. 

\begin{figure}
\includegraphics[width=\columnwidth]{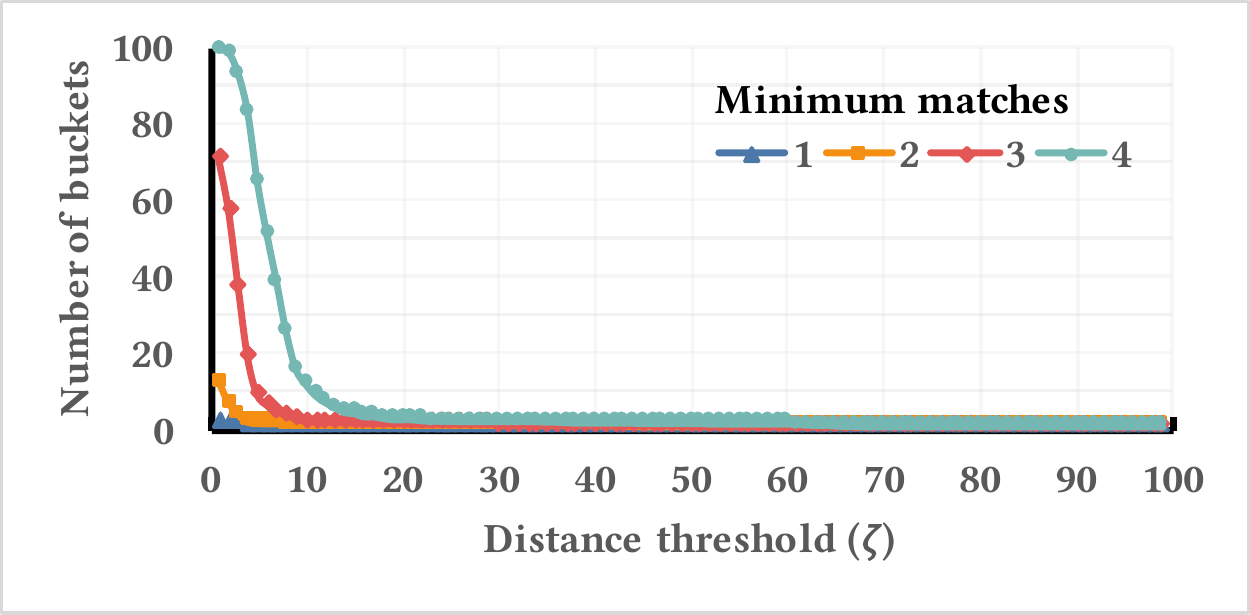}
\caption{Number of buckets as a function of distance threshold $\zeta$, with varying minimum matches.  \label{fig:repeat}}
\centering
\end{figure}

In Fig.~\ref{fig:repeat}, we observe how the average of the number of buckets for the seed documents changes with respect to the distance threshold $\zeta$ and a  \textit{minimum matches} score. $\zeta$ is varied from one to 100, since this is the range of the publication dates in this experiment, while the number of \textit{minimum matches} is changed from one to four.
Figure~\ref{fig:repeat} shows that as the distance threshold increases, the number of buckets rapidly decreases. This indicates that, even with different initializations of the optimization routine with multiple executions for the same seed, we obtain similar chains and time segments. This helps the user to keep the analysis focused on a relevant set of evolving concepts for the same seed document. 


\subsection{Prediction} \label{sec:exp:pred}
Our framework provides past documents as an evolution of the story of a given seed document. All the past documents have associated weights reflecting their importance in the evolution. 
These relevant documents can be used to study the evolution over time of the prominent entities found in the seed document. In this experiment, we examine if we can leverage these relevant documents to not only study the evolution of the entities in the past, but also to predict, using a simple linear regression, the future evolution. A more specific question is: which entities will appear in an article published in the future given a seed document?

For this experiment, we divided the set of relevant documents based on the segments into a training dataset (four segments) and a testing dataset (one segment). Next, we built a table which consists of all pairs of words in all pairs of the documents of the training set. This table contains the difference in publication dates, as well as a pair of entities and corresponding tf-idf weights that appear in a pair of documents. That is, the training data reflects how far in time a pair of entities may appear and what are the tf-idf scores of those entities when they appear. We built a linear regression model for each of the terms appearing in the \textit{future} documents, taking into account the difference in publication dates as an extra feature.
\begin{figure}
\includegraphics[scale=0.6]{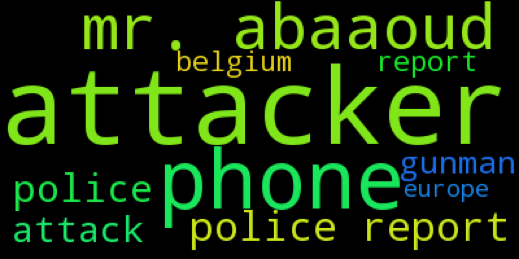}\\
(a)A word-cloud of a seed document.\\
\includegraphics[width=\columnwidth]{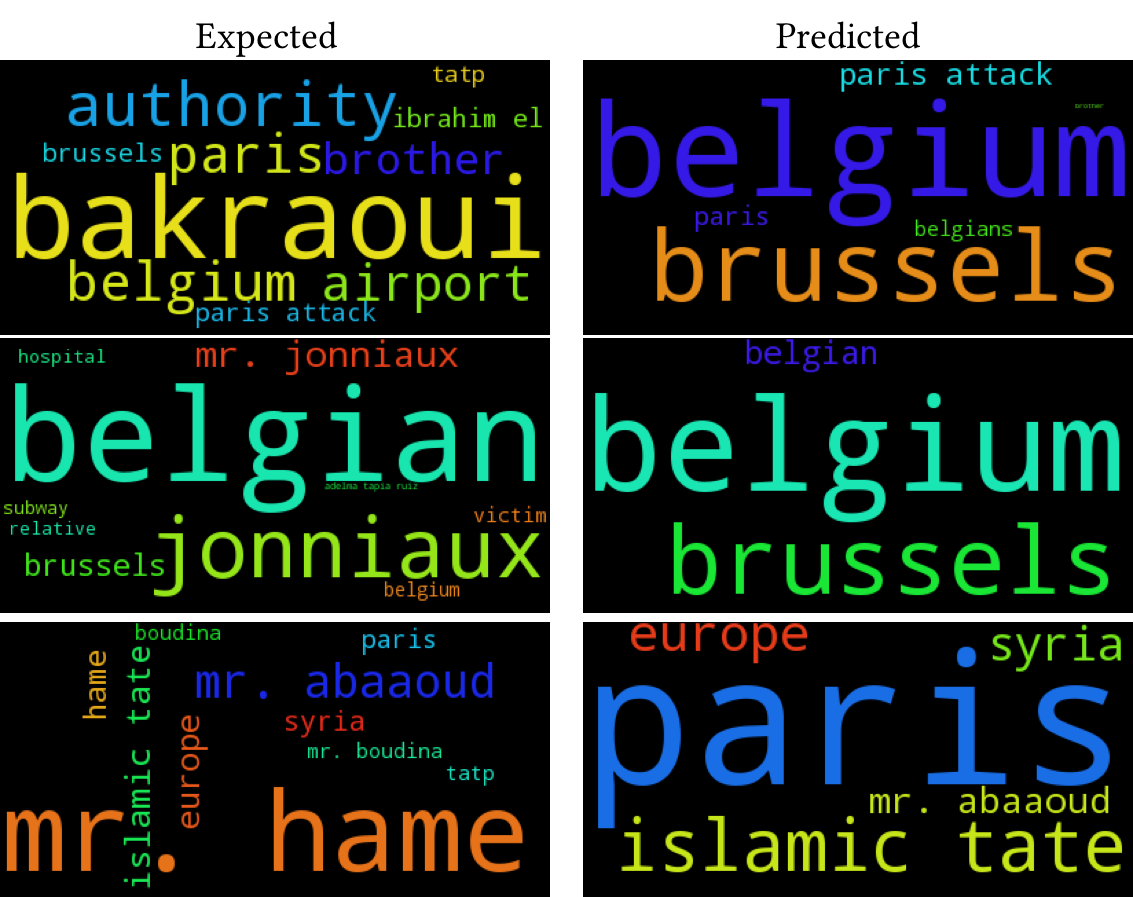}
(b) On the left, we show the ground truth while on the right we show the predicted word-clouds. The words are sized according to the original and predicted weights, respectively. The publication date difference between the seed document and the predicted documents ranges between 4 and 10 days.
\caption{An experiment on predicting a future document.
	\label{fig:wc_orig}}
\centering
\end{figure}

When a new prediction is requested, we use the linear regression models that include the entities of the seed document, to predict the weight of the entities in the \textit{future} document. In Fig.~\ref{fig:wc_orig}(a), we show the word-cloud of a sample seed document. We then use our algorithm to predict how the word-clouds of three future documents would look like in a range between 4 and 10 days after the publication date of the original article. The actual and predicted word-clouds are presented in Fig.~\ref{fig:wc_orig}(b). As can be seen, even with a simple linear regression-based prediction model, for most of the cases the predicted documents have a match of at least two of the entities from the ground truth (e.g., \textit{Paris attack}, \textit{Belgium}, and \textit{Paris}). The advantage of using the output of our model for prediction is that we exploit both the similarity of documents within a segment, and the diffusion of a topic across segments, to obtain better results. 

We expect that a more sophisticated algorithm would be able to improve these results. The main disadvantage of this method is that if no entities from the seed document appear in the training set, then no entities can be predicted, and of course, unseen entities will not be predicted.

\section{Conclusions and Future Work}
Our framework discovers the evolution of today's news stories from large news archives. We have presented case studies that demonstrate the possibility of using the resulting evolution-related documents as training data to predict the relevant entities for future news articles. Based on our case studies, we plan to expand our work in the direction of prediction where our framework will be able to generate warnings regarding insurgent activities, predict a timeline for developing a new concept, and forecast emerging stories. Furthermore, we plan to incorporate user feedback to improve the performance of our framework by modifying the optimization hyperparameters to meet the user's expectations. 



\bibliographystyle{ACM-Reference-Format}
\bibliography{kdd} 

\end{document}